\documentclass[twocolumn]{aastex631} 
\usepackage{graphicx}
\usepackage{multirow}
\usepackage[sort&compress]{natbib}
\usepackage{hyperref}

\defcitealias{2023ApJ...950...86G}{G23}
\defcitealias{1989ApJ...345..245C}{CCM89}
\defcitealias{2019ApJ...886..108F}{F19}

\newcommand\feh{{\rm [Fe/H]}}
\newcommand\logg{{\rm log}\,g}
\newcommand\teff{T_{\rm eff}}
\newcommand\ebv{E(B-V)}
\newcommand\sfdebv{E(B-V)_{SFD}}
\newcommand\gaia{Gaia}
\newcommand\RV{R_{\rm V}}

\shorttitle{XP extinction curve}
\shortauthors{Zhang et. al.}
\graphicspath{{./}{figures/}}

\begin{document}

\title{An Empirical Extinction Curve Revealed by Gaia XP Spectra and LAMOST}

\author[0000-0003-1863-1268]{Ruoyi Zhang} 
\affil{Institute for Frontiers in Astronomy and Astrophysics, Beijing Normal University, Beijing 102206, China; yuanhb@bnu.edu.cn}
\affil{Department of Astronomy, Beijing Normal University, No.19, Xinjiekouwai St, Haidian District, Beijing 100875, China}
\affil{Max Planck Institute for Astronomy, Königstuhl 17, Heidelberg D-69117, Germany}

\author[0000-0003-2471-2363]{Haibo Yuan} 
\author[0000-0002-1259-0517]{Bowen Huang} 
\author[0000-0002-4878-1227]{Tao Wang} 
\affil{Institute for Frontiers in Astronomy and Astrophysics, Beijing Normal University, Beijing 102206, China; yuanhb@bnu.edu.cn}
\affil{Department of Astronomy, Beijing Normal University, No.19, Xinjiekouwai St, Haidian District, Beijing 100875, China}

\author[0000-0002-9824-0461]{Lin Yang} 
\affil{Department of Cyber Security, Beijing Electronic Science and Technology Institute, Beijing 100070, China}

\author{Gregory M. Green}
\author[0000-0003-3112-3305]{Xiangyu Zhang} 
\affil{Max Planck Institute for Astronomy, Königstuhl 17, Heidelberg D-69117, Germany}

\begin{abstract}

We present a direct measurement of extinction curves using corrected $\gaia$ XP spectra \citep{2024ApJS..271...13H} of the common sources in $\gaia$ DR3 and LAMOST DR7.
Our analysis of approximately 370 thousand high-quality samples yielded a high-precision average extinction curve for the Milky Way.
After incorporating infrared photometric data from 2MASS and WISE, the extinction curve spans wavelengths from 0.336 to 4.6 $\mu$m.
We determine an average $R_{55}$ of $2.730 \pm 0.007$, corresponding to $\RV = 3.073 \pm 0.009$, and a near-infrared power-law index $\alpha$ of $1.935 \pm 0.037$.
Our study confirmed some intermediate-scale structures within the optical range.
Two new features were identified at 540 and 769\,nm, and their intensities exhibited a correlation with extinction and $\RV$.
This extinction curve can be used to investigate the characteristics of dust and enhance the extinction correction of Milky Way stars. 
A Python package for this extinction curve is available \citep{ruoyi_zhang_2024_12621834}.

\end{abstract}


\section{Introduction} \label{sec-intro}

The impact of dust on starlight in the universe is profound.
Dust grains absorb electromagnetic radiation across a wide spectrum, ranging from ultraviolet (UV) to infrared (IR) wavelengths, as well as X-ray wavelengths. 
Subsequently, they re-emit the absorbed energy in the form of infrared and microwave radiation \citep{2003ARA&A..41..241D}.
Re-radiation in the IR band could redistribute more than 30\% of the energy of starlight in the universe \citep{2002ApJ...571..107B}.
Moreover, dust grains scatter photons in the X-ray, UV, optical, and near-infrared bands.
The combined effect of these obscuration mechanisms on starlight, i.e., absorption and scattering, is known as dust extinction.
The variation of extinction with wavelength is the extinction curve or the extinction law.
Our understanding of interstellar dust largely relies on the study of extinction and its wavelength dependence.
The overall shape (particularly the slope) of the extinction curve provides crucial insights into the size distribution of dust grains, while its spectral features reflect the chemical composition of the dust.
Furthermore, the extinction curve serves as an indispensable correction tool for understanding the intrinsic properties of stars and galaxies.

OB stars are excellent probes of the extinction curve because their luminosity allows them to be observed out to large distances and through significant dust extinction.
Additionally, their spectra contain few spectral lines, thereby matched stellar templates are easy to find.
Consequently, the majority of extinction curve measurements to date have relied on the ``pair method" for OB stars.
Therefore, the majority of extinction curve measurements made so far have been conducted by using the ``pair method" for OB stars.
Pioneering studies by \citet{1986ApJ...307..286F,1988ApJ...328..734F} used UV spectra from the International Ultraviolet Explorer (IUE) satellite of 45 reddened OB stars and 10 standards to parameterize the UV extinction curve and the 2175\rm{\AA} bump for the first time. 
Subsequently, \citet[hereafter CCM89]{1989ApJ...345..245C} analyzed about 30 OB stars with IR photometry from \citet{1988ApJ...328..734F}.
They found that the total-to-selective extinction ratio, $\RV$, could account for most variations in the extinction curves from UV to optical wavelengths.
The \citetalias{1989ApJ...345..245C} extinction law was further validated by a much larger sample of 417 OB stars across diverse interstellar environments \citep{2004ApJ...616..912V}.
Subsequently, numerous studies have refined the Galactic extinction model using varied samples and wavelength ranges, employing diverse data analysis techniques \citep[e.g.,][]{1994ApJ...422..158O, 2009ApJ...705.1320G, 2014A&A...564A..63M, 2022ApJ...930...15D, 2023ApJ...950...86G}.
These models, dependent on $\RV$, have progressively revealed more detailed structures such as intermediate-scale structures (ISS).
Some study found ISSs are independent of $\RV$ \citep{2019ApJ...886..108F, 2020ApJ...891...67M}, 
while \citet{2023ApJ...950...86G} reached the opposite conclusion based on the same data.

These studies are carried out using multi-wavelength spectroscopic and photometric data from dozens to hundreds of OB stars. 
However, the relatively limited number of sightlines somewhat constrains our ability to extensively sample various interstellar environments. 
Consequently, it becomes challenging to conduct comprehensive statistical analyses that explore the relationships between the properties of dust, interstellar environments, and other aspects of the interstellar medium.

With the advent of large-scale sky surveys, measuring $\RV$ by combining spectroscopic and photometric data for a much larger sample of stars has overcome many limitations. 
\citet{2016ApJ...821...78S} utilized APOGEE spectra in conjunction with optical and near-infrared photometry from Pan-STARRS~1, 2MASS and WISE to map the $\RV$ distribution across 150,000 sightlines near the Galactic midplane.
They found that $\RV$ does not correlate with dust column density but exhibits a negative correlation with the far-infrared dust emissivity index.
\citet{2023ApJS..269....6Z} estimated $\RV$ values for approximately 3 million sightlines using datasets from LAMOST and multiple photometric sky surveys, thereby constructing a high-resolution 2D $\RV$ map that covers two-fifths of the sky.
They observed significant spatial correlations between $\RV$ and molecular clouds, while also verifying the general absence of correlation between $\RV$ and dust extinction. 
These investigations enhance our comprehension of dust characteristics in various interstellar conditions and provide accessible $\RV$ maps for accurate extinction corrections.

However, utilizing photometric data to calculate $\RV$ only offers an indirect assessment of dust properties, as it does not directly provide the extinction curve.
Additionally, extinction measurements derived from broadband photometry are impacted by bandwidth effects \citep[e.g.,][]{1956ApJ...123...64B, 1957ApJ...125..209B, 2023ApJS..264...14Z}, wherein the observed spectral energy distribution (SED) of stars influences the $\RV$ determinations.
Hence, there remains a need for large-scale extinction curve measurements to diminish selection biases and statistically explore the commonality of dust properties. 
This study leverages the most recent $\gaia$ XP (BP and RP) slitless spectral data, along with stellar parameters from LAMOST, to ascertain the extinction curve of the Milky Way.

This paper is structured as follows: 
Section \ref{sec-Data} outlines the data sources utilized in our analysis. 
In Section \ref{sec-Method}, we compute the intrinsic spectra and extinction curve using the star-pair method. 
Section \ref{sec-Result&Discussion} presents the calculation and validation of a median extinction curve and the identification of several intermediate-scale structures.
Our findings are summarized in Section \ref{sec-Summary}.

\section{Data} \label{sec-Data}
The $\gaia$ Data Release 3 \citep[hereafter DR3]{2023A&A...674A...1G} catalog was obtained by the first 34 months of continuous all-sky scans of the $\gaia$ mission of the European Space Agency. 
$\gaia$ DR3 has released mean low-resolution BP/RP spectra for approximately 220 million sources, representing the largest dataset of low-resolution spectra to date. 
The XP (BP and RP) spectra cover the wavelength range of 336 -- 1020\,nm with a spectral resolution of R $\sim$ 20 -- 70. 
\citet{2024ApJS..271...13H} corrected the systematic errors in the Gaia XP spectra by comparing them against external spectral libraries.
After comprehensively correcting the systematic errors depending on color $BP-RP$, $G$, and reddening, they achieved an internal precision of 1--2 percent.
We have incorporated their systematic error corrections in our analysis.

LAMOST DR7 accurately measures stellar parameters with the LAMOST Stellar Parameter Pipeline (LASP; \citealt{2011RAA....11..924W, 2015RAA....15.1095L}).
For our calculations, we utilized the effective temperature $\teff$, surface gravity $\logg$, and metallicity $\feh$ parameters from LAMOST.

\section{Method} \label{sec-Method}

\begin{figure*}
    \centering
    \includegraphics[width=\linewidth]{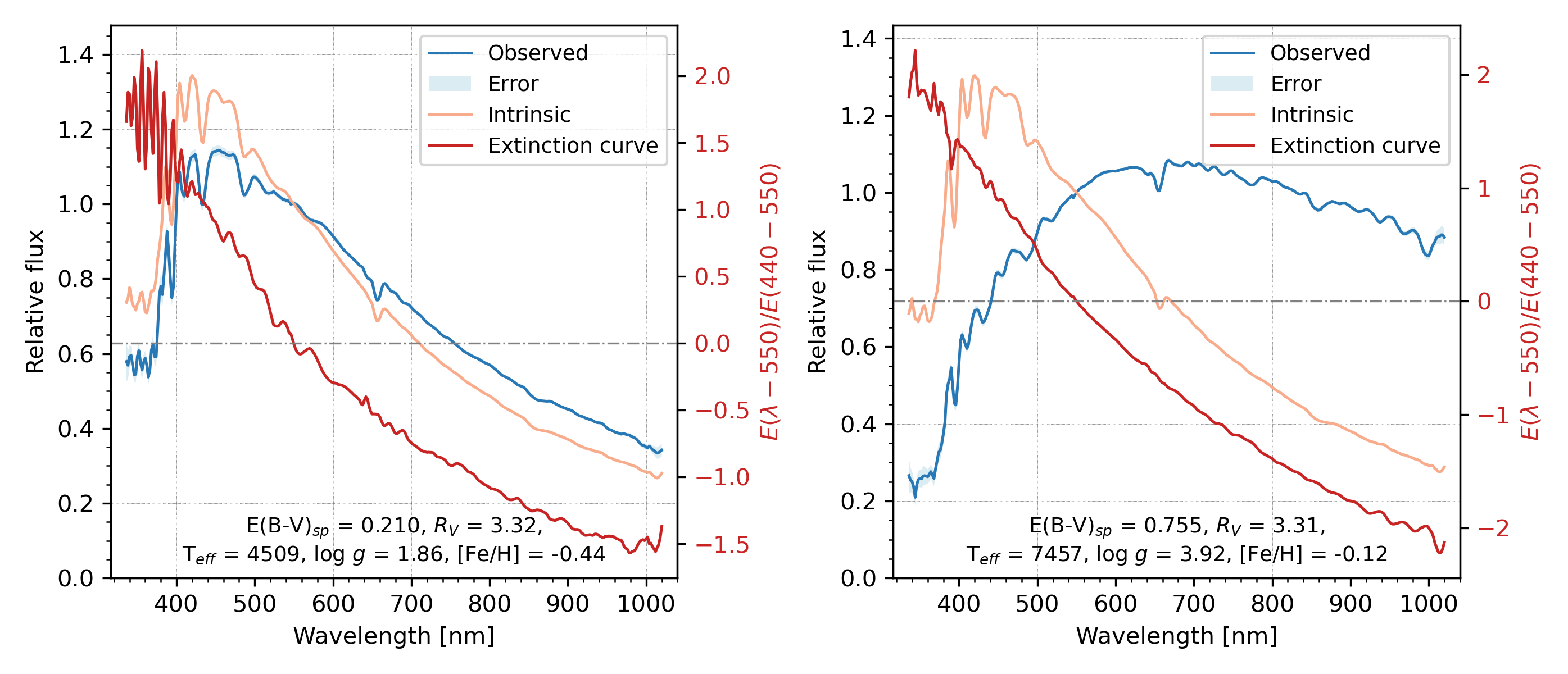}
    \caption{Examples showing the observed Gaia XP spectra (blue), the derived intrinsic Gaia XP spectra (orange), and the extinction curves (red) of two stars.
    Blue shading indicates errors in the observed spectrum.
    The left y-axis corresponds to the spectra and the right y-axis to the extinction curves.
    }
    \label{fig-spectra}
\end{figure*}

The star-pair algorithm, when applied to photometric survey data, can be used to obtain the intrinsic colors and reddening of stars based on the assumption that stars with the same stellar atmosphere parameters have the same intrinsic color \citep{1983ApJ...266..662M,2013MNRAS.430.2188Y,2023ApJS..264...14Z}. 
Similarly, we posit that stars with matching stellar parameters should also share identical intrinsic spectral profiles. 
By employing the star-pair method on XP spectra, we can deduce intrinsic spectra and extinction curves. 
For each reddened target star, the intrinsic spectrum is derived from its control pairs, also referred to as comparison pairs, which have similar atmospheric parameters and very low extinction. 
The extinction curve is then calculated by subtracting the intrinsic spectrum from the observed spectrum.

The target sample, i.e., the total sample, contains 5,056,929 common sources with XP spectra from Gaia and LAMOST. 
While the control sample stars are carefully selected based on the following criteria:
\begin{enumerate}
    \item 
    The vertical distance from Galactic disk $\vert$Z$\vert$ $>$ 300\,pc and the ecliptic latitude $\vert elat \vert$ $>$ 10$\degr$. 
    This ensures that the control sample remains unaffected by Galactic and potential zodiacal dust. 
    It is noteworthy that the Galactic dust disk's scale height is approximately 103\,pc \citep[e.g.,][]{2018ApJ...858...75L}.
    \item 
    The signal-to-noise ratio (SNR) of the LAMOST spectra is required to be greater than 20 to guarantee the accuracy of the derived stellar atmosphere parameters.
    \item 
    The spectrum's systematic error correction must be reliable \citep{2024ApJS..271...13H}.
    \item 
    $\sfdebv \le 0.01\,$mag; if $\feh < -0.8$, $\sfdebv < 0.01-(\feh+0.8)/100 $\,mag; if $\teff > 6300$\,K, $\sfdebv < 0.01+(\teff-6200)/(4\times10^4) $\,mag; if $\teff < 5500$\,K, $\sfdebv < 0.01-(\teff-5500)/(3\times10^5) $\,mag. We first adopted a stringent limit for the reddening, then relaxed the restriction in the low-temperature, high-temperature, and metal-poor zones, to obtain a sufficient number of control samples. At the high temperature end, the extinction limit is at a maximum of 0.1\,mag.
\end{enumerate}
Finally, we selected 67,650 (1.3\% of the total) control stars that uniformly cover various stellar parameters.

To mitigate random errors, we averaged the spectral flux within 542--558\,nm, designating the mean value as the flux at 550\,nm, denoted as $f_{550}$. 
Subsequently, for all observed spectra, we calculated the relative flux, $f_{\lambda}/f_{550}$, and dereddened the control sample therein.
Initially, we employed the extinction curve from \cite[][hereafter F19]{2019ApJ...886..108F} for the extinction correction. 
Once we acquired the median extinction curve based on the XP spectrum, it superseded the initial extinction curve for further correction.
This process is iterated until the results are stabilized.

For each star within the target sample, control pairs are determined from the entire control sample through a "pairing" process. 
A star is considered a suitable control pair if the differences in its parameters relative to those of the target star fulfill the following criteria: 
$\Delta\teff < 30 + [ 0.005 \times ( 6000 - \teff ) ]^2$\,K, 
$\Delta\logg < 0.5$\,dex, and 
$\Delta\feh < 0.3$\,dex.
Note that the threshold for temperature difference is minimized at 6000\,K and is gradually adjusted to be more lenient in regions where the control sample is less dense.
The maximum $\Delta\teff$ is about 340\,K at the highest temperature, 9500\,K.
The typical number of control stars assigned to each target star is approximately 800, and we set the minimum required number as ten.

To derive the intrinsic spectrum of the target star, we modeled the relationship between the relative flux and the stellar parameters of the local control pairs for each wavelength. 
This was accomplished using the function 
$f_{\lambda}/f_{550} = a \times \teff + b \times \logg + c \times \feh + d$, 
where $f_{\lambda}$ represents the dereddened flux at wavelength $\lambda$, 
and $a$, $b$, $c$, and $d$ are parameters to be determined. 
With these parameters established for each wavelength, the relative flux for any given star can be calculated by inputting its stellar parameters.

Given that the control sample is a subset of the target sample, the intrinsic fluxes estimated by the star-pair method are accessible, in addition to the intrinsic fluxes directly dereddened using the extinction curve. 
We then calculate the differences between these two sets of intrinsic fluxes at 400, 500, 650, 750, and 900 nm for the control sample, iteratively excluding 3$\sigma$ outliers. 

In this study, we use monochromatic normalization to mitigate the bandwidth effects in the extinction measurements.
Consequently, we adopt 440\,nm and 550\,nm as the approximations for the Johnson B and V bands, respectively \citep{2019ApJ...886..108F}.
To reduce the significant observational uncertainty in XP spectra,we average the extinction over a 16\,nm range centered at both 440\,nm and 550\,nm, thereby determining the monochromatic extinction.
The normalized extinction at wavelength $\lambda$, relative to that at 550\,nm can be calculated by
    \begin{equation} \label{eq-1}
        A_\lambda - A_{55} = -2.5\ log\frac{F^{'}_\mathrm{obs}(\lambda)}{F^{'}_\mathrm{intrinsic}(\lambda)},
    \end{equation}
where $A_\lambda$ and $A_{55}$ is the extinction at wavelength $\lambda$ and 550\,nm, 
$F^{'}(\lambda)$, defined as $F(\lambda)/F(55)$, represents the flux at $\lambda$ relative to the flux at 550\,nm, the subscripts ``$\mathrm{obs}$" and ``$\mathrm{intrinsic}$" respectively denote the observed and intrinsic SED, respectively.
We define the normalized extinction curve, in alignment with \citetalias{2019ApJ...886..108F}, as:
    \begin{equation} \label{eq-2}
        k(\lambda-55) \equiv \frac{A_\lambda - A_{55}}{A_{44} - A_{55}}
    \end{equation}
and the definition of the ratio of total-to-selective extinction $R_{55}$ as 
    \begin{equation} \label{eq-3}
        R_{55} \equiv \frac{A_{55}}{A_{44} - A_{55}} = \frac{A_{55}}{E(44-55)}.
    \end{equation}
Here, $k(\lambda-55)$ and $R_{55}$ serve as analogs to the more commonly used extinction normalizations, $k(\lambda-V)$ and $\RV$, respectively.

Figure\,\ref{fig-spectra} illustrates the calibrated observed spectra, the intrinsic spectra, and the extinction curves for a giant star and a main-sequence star. 
The observable fluctuations, or "wiggles", present in the observed spectra and the extinction curves are primarily the artifacts in the XP spectra due to long-range noise correlations \citep{2023A&A...674A...2D}. 
The intrinsic spectrum for each star is estimated using multiple local control pairs, allowing their wiggles to counterbalance each other. 
Finally, we obtained the extinction curves for 99.5\% of the target sample.

\section{Result and Discussion} \label{sec-Result&Discussion}

\subsection{The Extinction curve} \label{sec-extinction_curve}

\begin{figure*}
    \centering
    \includegraphics[width=\linewidth]{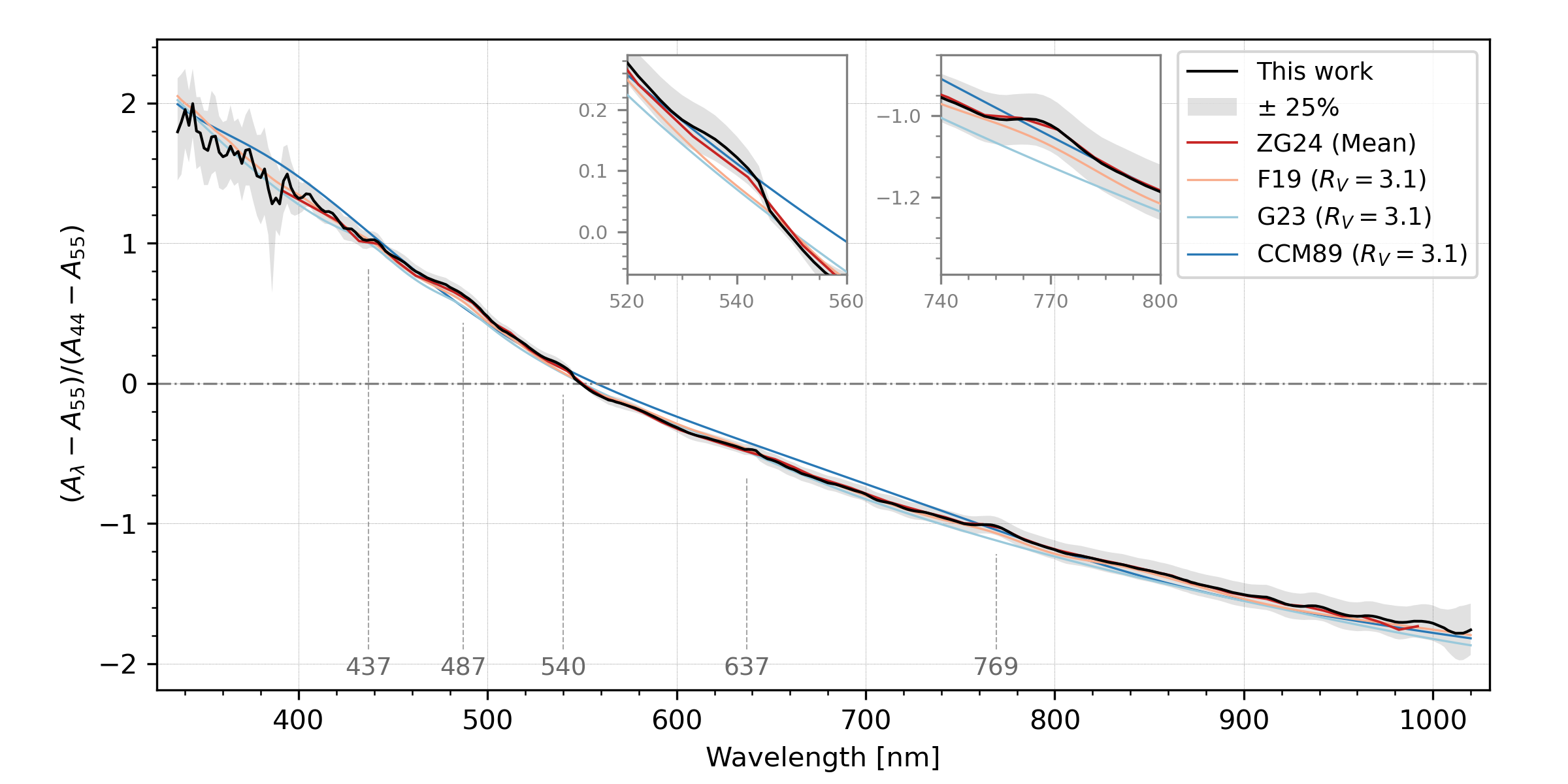}
    \caption{The median extinction curve and its comparison with literature.
    The black line is the median extinction curve, while the gray shadow indicates the distribution range for 25\% to 75\% of stars. 
    The models from ZG24, \citetalias{2019ApJ...886..108F}, \citetalias{2023ApJ...950...86G}, and \citetalias{1989ApJ...345..245C} are illustrated by red, pink, light blue, and blue lines, respectively.
    The gray dashed lines and the wavelength labels indicate the locations of the possible ISSs.}
    \label{fig-MEC}
\end{figure*}

We selected a high-quality sample of 366,189 stars to derive the median extinction curve by the following selection criteria: 
(1) $\ebv$\footnote{Calculated using the color excess and reddening coefficient of color $G_{\rm BP}-G_{\rm RP}$ in \citet{2023ApJS..264...14Z}. The same applies hereinafter.} $>$ 0.15\,mag; 
(2) LAMOST SNR $>$ 30;
(3) G $>$ 12\,mag;
(4) Reliable systematic error correction \citep{2024ApJS..271...13H}. 
Afterward, we removed the 3$\sigma$ outliers of $k(\lambda-55)$ at each wavelength and calculated the median $k(\lambda-55)$, as displayed in Figure\,\ref{fig-MEC}.
Note that the small wiggles at both the blue and red of the extinction curve are artifacts in the XP spectra.
And the spike around 641\,nm lies close to the boundary between the BP and RP bands at 635\,nm, suggesting that it could be artificial.

In Figure\,\ref{fig-MEC}, we demonstrate the mean extinction curve with $\RV=3.07$ from ZG24, and the model curve with $\RV=3.1$ of \citetalias{2019ApJ...886..108F}, \citet[][hereafter G23]{2023ApJ...950...86G}, and \citet[][hereafter CCM89]{1989ApJ...345..245C} for comparison.
Overall, the consistency of the extinction curves is good.
The differences between the ZG24 and XP extinction curves are less than one-percent, while the differences for \citetalias{2019ApJ...886..108F} and \citetalias{2023ApJ...950...86G} are within 5\%. The \citetalias{1989ApJ...345..245C} extinction curve differs by up to 10\% in the 540-760\,nm range and below 440\,nm. 
More detailed comparison will discuss in Section \ref{sec-literature}.

\begin{figure*}
    \centering
    \includegraphics[width=\linewidth]{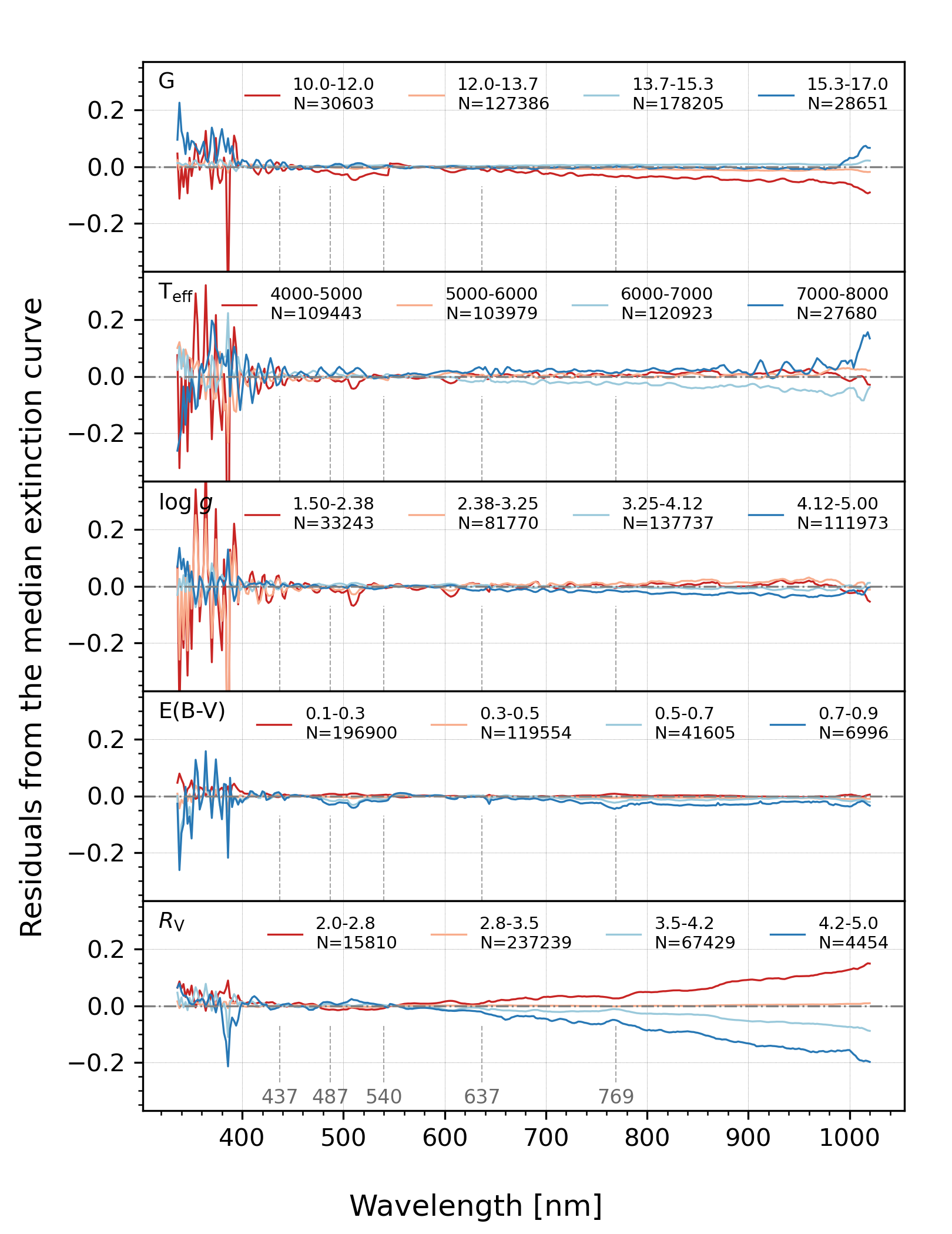}
    \caption{The difference between the average extinction curve of samples within varying parameter ranges and the overall median curve.
    From top to bottom, these panels display the impact of the following parameters: G band magnitude, $\teff$, $\logg$, $\ebv$, and $\RV$. 
    Colors represent different ranges of each parameter.
    }
    \label{fig-relevance}
\end{figure*}

To validate our extinction curve, we assessed its correlation with several stellar or ISM parameters, including the G band magnitude, $\teff$, $\logg$, $\ebv$, and $\RV$.
We divided the high-quality sample into four subsets on each parameter\footnote{For the G magnitude, we included stars of $10 < G < 12$.} and calculated their respective median extinction curves.
Subsequently, we analyzed the residuals between these extinction curves and the extinction curve of the entire high-quality sample.
The results are shown in  Figure\,\ref{fig-relevance}.

The curves in the top panel agree well, except for the red line which represents stars with $10<G<12$, for which new systematic errors were introduced during the calibration \citep{2024ApJS..271...13H}.
Consequently, stars of $10<G<12$ has been excluded from the high-quality sample.
The curves exhibit little correlation with both $\teff$ and $\logg$, suggesting that these two critical stellar parameters in the star-pair method do not introduce significant systematic errors in our results. A slight correlation with $\ebv$ is observed.
This could be attributed to the distinct spatial distributions of the subsets.
Variations in dust properties, associated with different spatial positions, might alter the shape of the extinction curve \citep{2023ApJS..269....6Z}.
A secondary reason might be that not all XP systematic errors related to $\ebv$ have been fully corrected, particularly in the high extinction end. 
The $\RV$ data are from measurements labeled as ``reliable" in \cite{2023ApJS..269....6Z}, which employs forward modeling along with a series of photometric data spanning from ultraviolet to infrared.
Hence, the correlation between $\RV$ and the extinction curves is expected.

\subsection{Comparison with photometric reddening} \label{sec-photo_reddening}

Observed color excess ratios provide an independent method for testing extinction curves. 
Therefore, we use precise reddening measurements for 5 million stars from \citet{2023ApJS..264...14Z} to further examine the extinction curve.
To integrate photometric and spectroscopic data, this study assumes two ultra-narrow passbands of width 16\,nm to calculate the magnitudes at 440 and 550\,nm, respectively. 
We used the star-pair algorithm to derive the color excess for colors (440-550) and (550-Ks).
Consequently, we can represent the empirical reddening for each color as CERs in the form of $k(\lambda-55)$.
On the other hand, we adopt a forward modeling approach to calculate simulated CERs \cite[]{2023ApJS..269....6Z}, using
the median XP extinction curve, the BOSZ synthetic spectral database \citep{2017AJ....153..234B}, and the response function for each filter \citep{2020sea..confE.182R}.

\begin{figure*}
    \centering
    \includegraphics[width=\linewidth]{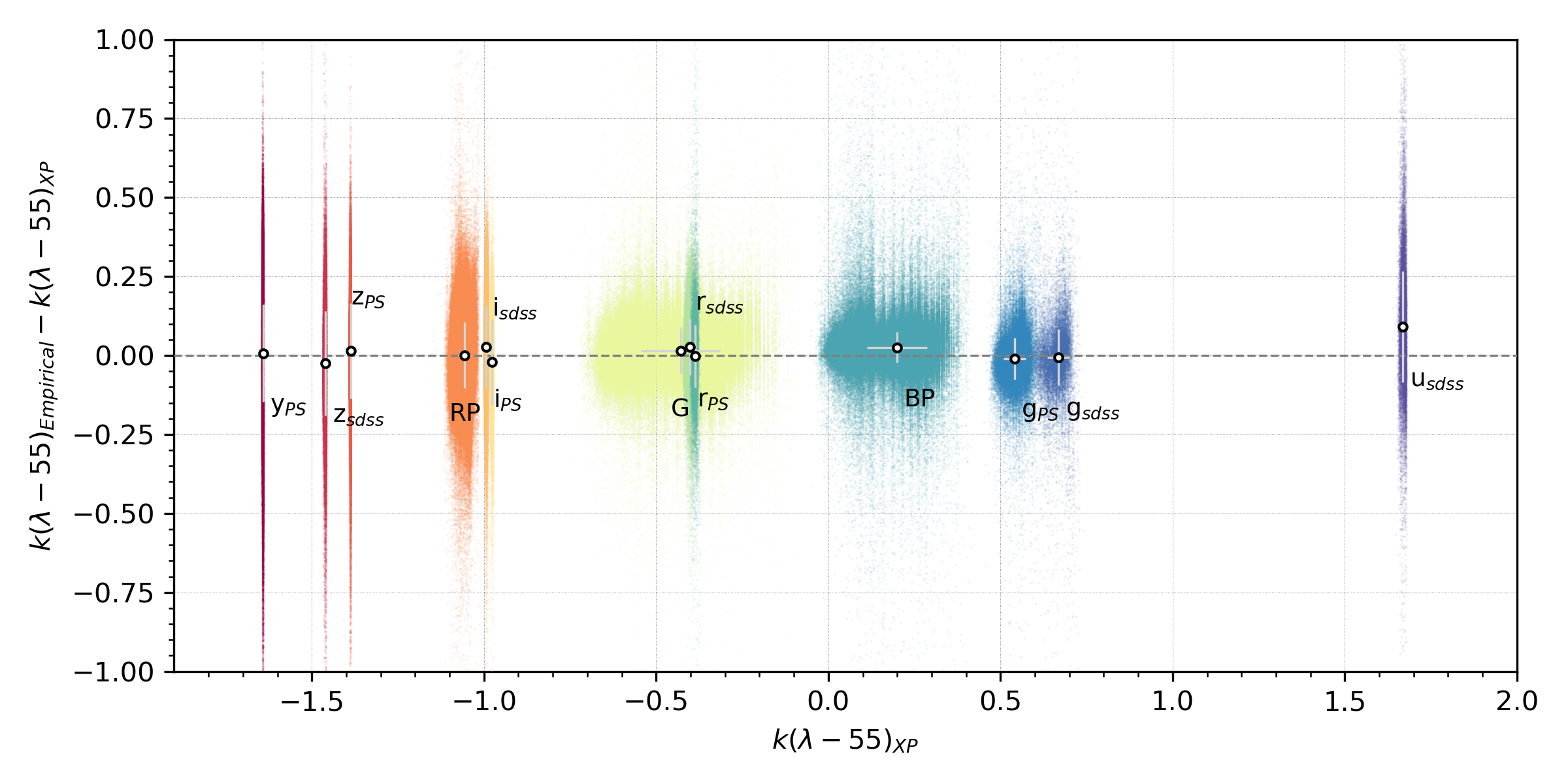}
    \caption{The comparison between observed and simulated color excess ratios (CERs) for the high-quality sample. The Y-axis represents the observed photometric CERs, and the X-axis is the simulated CERs, derived from the median XP extinction curve and stellar atmospheric model. The color coding indicates the specific photometric band, with a label on the side.  Black open circles are medians for the data points within the same band, with the error bar denoting the standard deviation. 
    }
    \label{fig-Phot}
\end{figure*}

Figure\,\ref{fig-Phot} compares the observed and simulated $k(\lambda-55)$ of the high-quality sample.
The dispersion of observed CERs arises mainly from random errors in the magnitudes of the 440 and 550\,nm bands, which depend on the signal-to-noise ratio of the XP spectra.
The dispersion in the simulated CERs merely reflects variations in SED.
The strong agreement between the two sets of CERs confirms the reliability of our results, particularly considering that they share the same dependence on SED. 

\subsection{Intermediate-scale structures} 
\label{sec-ISS}
It is noteworthy that, besides the wiggles at both ends, there are some distinct real structures in the extinction curve. 
We suspect that the bumps at 487, 540, 637, and 769\,nm, respectively, are the so-called intermediate-scale structures (ISS).
ISSs are usually broader than diffuse interstellar bands (DIBs), but narrower than the commonly observed broadband variability, as reported in \citet{1966ApJ...144..305W,2020ApJ...891...67M}.

Using extinction curves of 72 early-type stars from \citetalias{2019ApJ...886..108F}, \cite{2020ApJ...891...67M} identified three optical ISSs at 437, 487, and 630\,nm, respectively.
Our study also observed bumps near 487 and 630\,nm. 
However, the central wavelengths for the feature at 630\,nm are shifted to 637\,nm, possibly due to the influence of the junction of the BP and RP spectra.
The feature at 487\,nm is very distinct and can be seen from all sources except \citetalias{1989ApJ...345..245C}.
Whereas the 437\,nm ISS is located near the blue end where the wiggles are strong, thus affects its profile.

The newly found candidates for ISS at 540 and 769\,nm are evident in this study and in ZG24, both using XP spectra.
ZG24 employed a forward-modeling machine learning approach to obtain the intrinsic spectra and extinction curves for sources common to both $\gaia$ XP and LAMOST DR7. 
Compared to the three known ISSs, these two new features exhibit narrower widths and weaker intensities. 
As a result, they were not prominent in previous literature extinction curves, though still can be observed. 
For example, a noticeable bump at 769\,nm (1.30\,$\mu m^{-1}$) and a slight bump within a trough at 540\,nm (1.85\,$\mu m^{-1}$) are visible in Figure\,2 from \cite{2020ApJ...891...67M} and the second panel of Figure\,5 from \citetalias{2023ApJ...950...86G}. 

The extinction models of \citetalias{1989ApJ...345..245C} and \citetalias{2023ApJ...950...86G} did not exhibit these structures, since they both employed low order polynomial to fit the continuum.
Such fitting techniques smoothed out the finer details in the extinction curve, potentially obscuring subtle structures.

In this study, there are three ISSs that can be clearly detected, namely at 487, 540, and 769\,nm.
As depicted in Figure\,\ref{fig-relevance}, the intensity of these three ISSs does not vary with the physical parameters of stars, further supporting their reality.
It is noteworthy that they all have a negative correlation with $\ebv$.
This work, in contrast to the \citetalias{2019ApJ...886..108F} samples used in previous studies, employed more stars with lower extinction, resulting in these two ISSs being detected with greater intensity.

Among these three ISSs, only the one at 769\,nm shows a significant correlation with $\RV$, whereas the features at 487 and 540\,nm do not.
\cite{2020ApJ...891...67M} found that the intensities of the three ISSs located at 437, 487, and 630\,nm are independent of $\RV$.
Conversely, using the same data, \citetalias{2023ApJ...950...86G} observed a correlation with $\RV$. 
However, \citetalias{2023ApJ...950...86G} did not find $\RV$ relevance for ISSs at 540 and 769\,nm.
These conflicting results indicate that the correlation of these ISSs with $\RV$ requires further study, which is beyond the scope of this paper.
Our Future research will explore the potential connections of ISSs with interstellar dust, DIBs, and gas.

\subsection{Extend to the infrared}
\label{sec-ExtendCurve}
\begin{figure*}
    \centering
    \includegraphics[width=\linewidth]{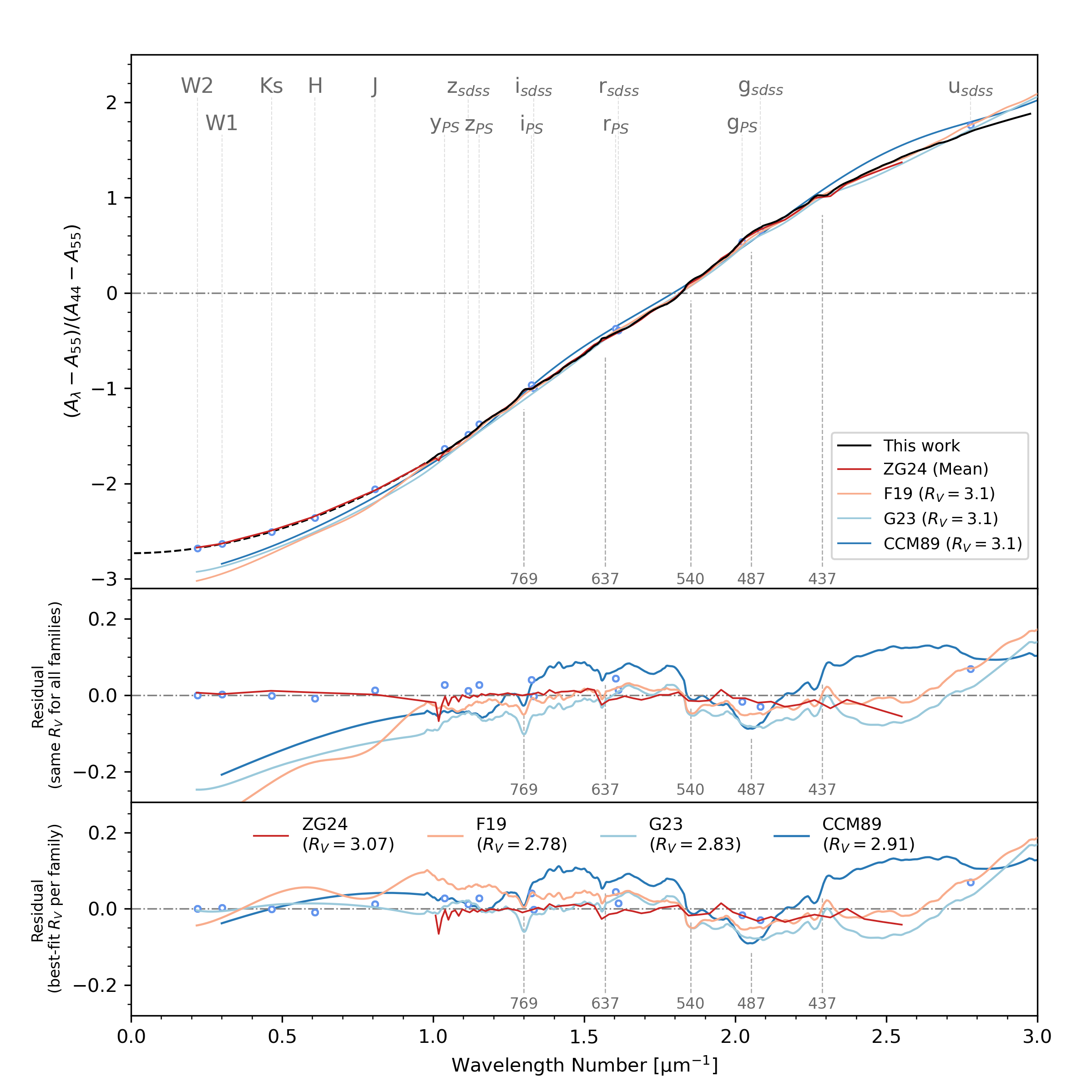}
    \caption{
    Comparisons between the median extinction curve and literature models along with photometric data.
    {\it Top panel}: The black solid and dashed lines are the median extinction and infrared extension curves, respectively. The red, pink, light blue, and blue lines are extinction models from ZG24, \citetalias{2019ApJ...886..108F}, \citetalias{2023ApJ...950...86G}, and \citetalias{1989ApJ...345..245C}, respectively. Blue open circles indicate the median $k(\lambda-55)$ and median $\lambda_{\rm{eff}}$ for each passband, with error bars included, although they are nearly imperceptible. The names of the corresponding passbands are labeled at the top. The gray dashed lines and the wavelength labels indicate the locations of the possible ISSs.
    {\it Middle panel}: The residuals obtained by subtracting the median extinction curve from the four models in the top panel.
    {\it Bottom panel}: The residual curves corresponding to the best-fit $\RV$ values.
    }
    \label{fig-liter}
\end{figure*}

Using the CERs described in Section \ref{sec-photo_reddening}, we can integrate the photometric data, especially the infrared bands from 2MASS and WISE, with our extinction curve.
First, we calculate the effective wavelength of passbands using the following function:
    \begin{equation} \label{eq-6}
        \lambda_{\rm{eff}} = \frac{\int \lambda^2 T(\lambda) F(\lambda) d\lambda}{\int \lambda T(\lambda) F(\lambda) d\lambda} ,
    \end{equation}
where $T(\lambda)$ is the filter transmission. $F(\lambda)$ is the BOSZ synthetic spectra for corresponding stellar parameters of each star, which is reddened using the XP extinction curve at specific $\ebv$.
Figure\,\ref{fig-liter} displays the effective wavelengths and median CERs for each band of the high-quality sample.
Consistent with the previous comparisons, the two types of extinction data match very well.

To mitigate wiggles at both ends of the XP median extinction curve, we then smoothed the extinction curve for $\lambda < 440$\,nm and $\lambda > 1000$\,nm. 
Subsequently, we perform a power-law fit for the smoothed extinction curve at wavelengths between 1000\,nm and 1020\,nm, as well as for the $k(\lambda-55)$ values in the J, H, Ks, W1, and W2 bands. 
The power-law function is described by:
    \begin{equation} \label{eq-7}
        k(x-55) = b \cdot x^{\alpha} - R_{55} ,
    \end{equation}
where $x$ is the wavelength number in $\mu m^{-1}$, $b$ is a free parameter, $\alpha$ is the power-law index, and $R_{55}$ represents the intercept at $x=0$.

Our fitting yield $ b = 0.994 \pm 0.007 $, $ \alpha = 1.935 \pm 0.037 $, and $R_{55} = 2.730 \pm 0.007 $.
By adopting $\alpha = 0.910$ and $\beta = 0.064$ from ZG24 for Eq.\,\ref{eq-5}, we convert $R_{55}$ into $\RV = 3.073 \pm 0.009$.

By combining the smoothed median extinction curve with the power-law function for the infrared region, we can produce an extended extinction curve spans from 0.336 to 4.6 $\mu$m.
This extended extinction curve is suitable for extinction corrections across various stellar spectra in the Milky Way.

\subsection{Comparison with literature}
\label{sec-literature}
 To compare with the curves from from ZG24, \citetalias{2019ApJ...886..108F}, \citetalias{2023ApJ...950...86G}, and \citetalias{1989ApJ...345..245C}, we first need to convert them all into the form of $k(\lambda-55)$.
ZG24 and \citetalias{2019ApJ...886..108F} do not require conversion, whereas \citetalias{2023ApJ...950...86G} and \citetalias{1989ApJ...345..245C} use the following formula for conversion:
\begin{equation} \label{eq-4}
        k(\lambda-55) = \alpha\ k(\lambda-V) + \beta
    \end{equation}
and
    \begin{equation} \label{eq-5}
        R_{55} = \alpha\ \RV - \beta ,
    \end{equation}
where $\alpha=0.990$ and $\beta=0.049$ (Eq.\,16,\,17 of \citetalias{2019ApJ...886..108F})
We compared them with our median extinction curve as illustrated in Figure\,\ref{fig-liter}.
In the visible range (400-700\,nm), the literature curves generally align well with ours.
The differences for \citetalias{2019ApJ...886..108F} and \citetalias{2023ApJ...950...86G} remain within 5\%. While the \citetalias{1989ApJ...345..245C} extinction curve shows up to a 10\% difference in the 540-760\,nm range.
Using similar data, the average extinction curve of ZG24 displays high consistency with our findings, including the ISSs mentioned above.
The overall residuals are less than one percent.
The relatively larger discrepancies within the 400-500\,nm and 625-645\,nm ranges are probably attributed to the systematic errors in the XP spectra within these bands \citep{2024ApJS..271...13H}, which have been addressed in our corrections. 
This mutual validation further attests to the validity of the results from both studies.

Except for ZG24, all models at $\RV=3.1$ exhibit significant discrepancies with the photometric $k(\lambda-55)$ in the infrared region with a wavelength larger than 950\,nm. 
This finding aligns with that of \cite{2019ApJ...877..116W}, who reported a power-law index $\alpha$ of $2.070 \pm 0.030$.
The infrared extinction data used in this paper have been verified to be consistent with their color excess ratios \citep{2023ApJS..264...14Z}.
For reference, the infrared spectrum from 0.8 to 4.0 µm used in \citetalias{2023ApJ...950...86G} is sourced from \cite{2022ApJ...930...15D}, with an average $\alpha$ of 1.7.
A possible explanation of this discrepancy is the selection effect of the sample, meaning that the power-law index $\alpha$ of the infrared extinction curve varies with the line of sight \citep[e.g.,][]{2009ApJ...696.1407N,2011ApJ...737...73F,2017ApJ...849L..13A,2022ApJ...930...15D}.

\section{Summary} \label{sec-Summary}

We have directly measured the extinction curves using $\gaia$ XP spectra of approximately five million common sources from $\gaia$ DR3 and LAMOST DR7. 
We derived the average extinction curve for the Milky Way from about 370 thousand high-quality spectra. 
The wavelength range of this curve includes two segments: 
spectroscopic coverage from 336 to 1020\,nm, and an extension to 4.6\,$\mu$m through power-law fitting to the infrared photometric data. 
Validation against photometric data and existing extinction models confirms the reliability of our extinction curve. 

The median extinction curve reveals several optical structures with resolution down to approximately 10\,nm. 
The ISS features at 487 and 630\,nm are confirmed, with new features identified at 540 and 769\,nm. 
For the high-quality sample, whose $\ebv$ is greater than 0.15, the average $ R_{55}$ is $ 2.730 \pm 0.007 $, corresponding to $\RV = 3.073 \pm 0.009$.
The near-infrared power-law index $ \alpha $ is $ 1.935 \pm 0.037 $.
This direct measurement of extinction curves, utilizing the largest spectral dataset to date, provides a powerful tool for studying dust properties and correcting extinction in stellar spectra across the Milky Way.

\begin{acknowledgments}

This work is supported by the National Key Basic R\&D Program of China via 2019YFA0405500 and the National Natural Science Foundation of China through the projects NSFC 12222301, 12173007, and 12173034.
We acknowledge the science research grants from the China Manned Space Project with NO. CMS-CSST-2021-A08 and CMS-CSST-2021-A09. 

This work has made use of data products from the LAMOST, $\gaia$,  2MASS, and WISE. 
Guoshoujing Telescope (the Large Sky Area Multi-Object Fiber Spectroscopic Telescope; LAMOST) is a National Major Scientific Project built by the Chinese Academy of Sciences. 
Funding for the project has been provided by the National Development and Reform Commission. 
LAMOST is operated and managed by the National Astronomical Observatories, Chinese Academy of Sciences. 

\end{acknowledgments}

%



\clearpage
\bibliography{ref.bib}



\end{document}